\documentclass[11pt, twocolumn]{article}
\usepackage[affil-it]{authblk}
\usepackage{blindtext}
\usepackage{abstract}
\usepackage{amsmath,amssymb,amsfonts,latexsym, mathtools}
\usepackage[T1]{fontenc} 
\usepackage[utf8]{inputenc} 
\usepackage{graphicx} 
\usepackage[margin=10pt,font=small,labelfont=bf, hypcap=false]{caption} 
\usepackage{subcaption}
\usepackage[usenames, dvipsnames]{color}
\usepackage{color} 
\usepackage{siunitx} 
\usepackage{paracol}
\usepackage{float}
\usepackage{hyperref}
\usepackage{todonotes}

\newcommand{\subfigimg}[3][,]{%
  \setbox1=\hbox{\includegraphics[#1]{#3}}
  \leavevmode\rlap{\usebox1}
  \rlap{\hspace*{40pt}\raisebox{\dimexpr\ht1-5\baselineskip}{#2}}
  \phantom{\usebox1}
} 
\newcolumntype{C}[1]{>{\centering\let\newline\\\arraybackslash\hspace{0pt}}m{#1}}
\makeatletter
\let\ftype@table\ftype@figure
\makeatother
\begin{document}

\title{\Huge Experimental study of elastic properties of noodles and bucatini}
\author[$\dagger$]{Vladimir Vargas-Calderón}
\author[$\dagger$]{A. F. Guerrero-González}
\author[$\dagger$]{F. Fajardo}
\affil[$\dagger$]{Departamento de Física - Universidad Nacional de Colombia \\Bogotá, Colombia}

\date{}
\twocolumn[
\begin{@twocolumnfalse}
\maketitle

\begin{abstract}
\noindent The mechanical behaviour of two types of pasta (noodles and bucatini) was studied in a cantilever-loaded-at-the-end experimental setup. One end of each pasta was fixed while the other end was submitted to forces perpendicular to the line determined by the pasta when undeflected.  Elastic curves were studied, resulting in values of $E=2.60\ \si{\giga\pascal}$ and $E=2.26\ \si{\giga\pascal}$ for the Young's modulus of bucatini and noodles respectively. The relation coming from small slopes approximation between the free end's displacement and the load was analyzed, resulting in values of $E=2.33\ \si{\giga\pascal}$ and $E=2.44\ \si{\giga\pascal}$ for the Young's modulus of bucatini and noodles respectively. Mechanical hysteresis was found in the pasta, resulting in a small deformation. This experiment can be done with low cost materials and it is a good first introduction to some basic concepts of elasticity for mechanics courses.\\\\
E-mail: {\tt vvargasc@unal.edu.co}
\end{abstract}
\vspace{0.3cm}
\noindent\rule{19cm}{0.5pt}
\vspace{0.5cm}
\end{@twocolumnfalse}
]
\saythanks
\section{Introduction}
Elasticity theory has been widely developed to study the behaviour of several materials submitted to forces. This theory has multiple applications in engineering and physics. In particular, small deflections of objects like beams when external forces act upon them have been deeply studied \cite{Landau, Fajardo}.

The study of materials' behaviour submitted to forces is of paramount importance in areas such as structural analysis, through which it can be determined the load bearing capacity of a structure, and to compute the dimensions and the materials to be used for its construction. This is exemplified in many studies, like the ones carried by Gonzáles  \textit{et al.} \cite{Guadua}, Contreras \textit{et al.} \cite{Pino_1} and Astori \textit{et al.} \cite{Pino_2}, where beams laminated with bamboo and pine were studied to measure their elastic properties with the goal of obtaining data about their resistance, their structural quality, among others, in order to be used as construction materials.

With regard to edible pasta (spaghetti, bucatini, noodles, among others) it is normal to apply this theory to make the products undergo quality controls. The main goal is to determine how the environment conditions like humidity, pressure and temperature affect the rheological properties of pasta, in particular the rigidity \cite{rheology_broccoli}. Notwithstanding, there does not exist a global standard of specific concentrations of the pastas' ingredients, allowing its composition to be varied from factory to factory. This fact is clearly reflected, for instance, in the variation of the Young's modulus reported for pasta from different parts of the world. Some Young's modulus of spaghetti were measured to be 5 \si{\giga\pascal} \cite{brittle}, 4.75 \si{\giga\pascal} \cite{iran}, 0.474-0.610 \si{\giga\pascal} \cite{sweet_potato}, 0.05-0.20 \si{\giga\pascal} \cite{fried_noodle}. It is important to note that different measurement methodologies were used in these studies.

In the context of our work, elastic behaviour of a variety of pasta (noodles and bucatini) with different geometries (elliptical and ring-like cross section, respectively) was analysed. The pastas were set up as cantilever beams, submitted to a force in the free end. In the Theory section of the document, the small slopes approximation is summarily exposed, the equation that models the curvature generated on the pastas when they are submitted to a force is deduced, and the relation between the vertical displacement of the free end and the force is showed. In the Experimental Setup section, the physical characteristics of the pasta used is described, as well as the materials used in the setup, and the cares taken to perform the measurements. In the Results and Analysis section, the Young's modulus measurements, elastic curves and hysteresis curves of the pasta are presented. The main conclusions are shown in the last section.

The experiment carried out is of low cost, and allows physics and engineering first-year students to approach the qualitative and quantitative study of the basic aspects of the material's theory of elasticity.

\section{Theoretical Framework}
\begin{figure}[h!]
    \centering
    \includegraphics[width=0.4\textwidth]{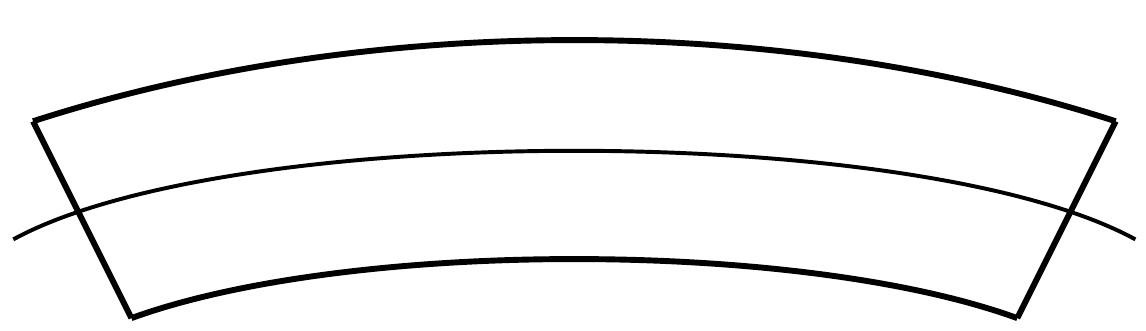}
        \caption{Neutral axis of a bent beam.}\label{línea neutra}
\end{figure}

A beam is a solid which, when free from external forces, is symmetric with respect to a given straight axis \cite{Amigo_de_fajardo}. This solid can be thought as the junction of many thin sheets, each of them perpendicular to the straight axis. These sheets are called cross sections. If all the cross sections are identical then the beam is characterized by a uniform cross section. The solid can also be thought as the junction of several thin fibers parallel to the mentioned axis. If the beam is bent because of an external force, then the fibers on the upper part of the beam will be stretched, while the fibers on the lower part of the beam will be shrunk, as it can be seen from Figure~\ref{línea neutra}. However, there is a fiber that is neither stretched nor shrunk. If the beam has a uniform cross section this fiber passes through each cross section's center of mass and it is the same as the straight axis, called the neutral axis \cite{Timoshenko}.

\begin{figure}[ht!]
    \centering
    \includegraphics[width=0.45\textwidth]{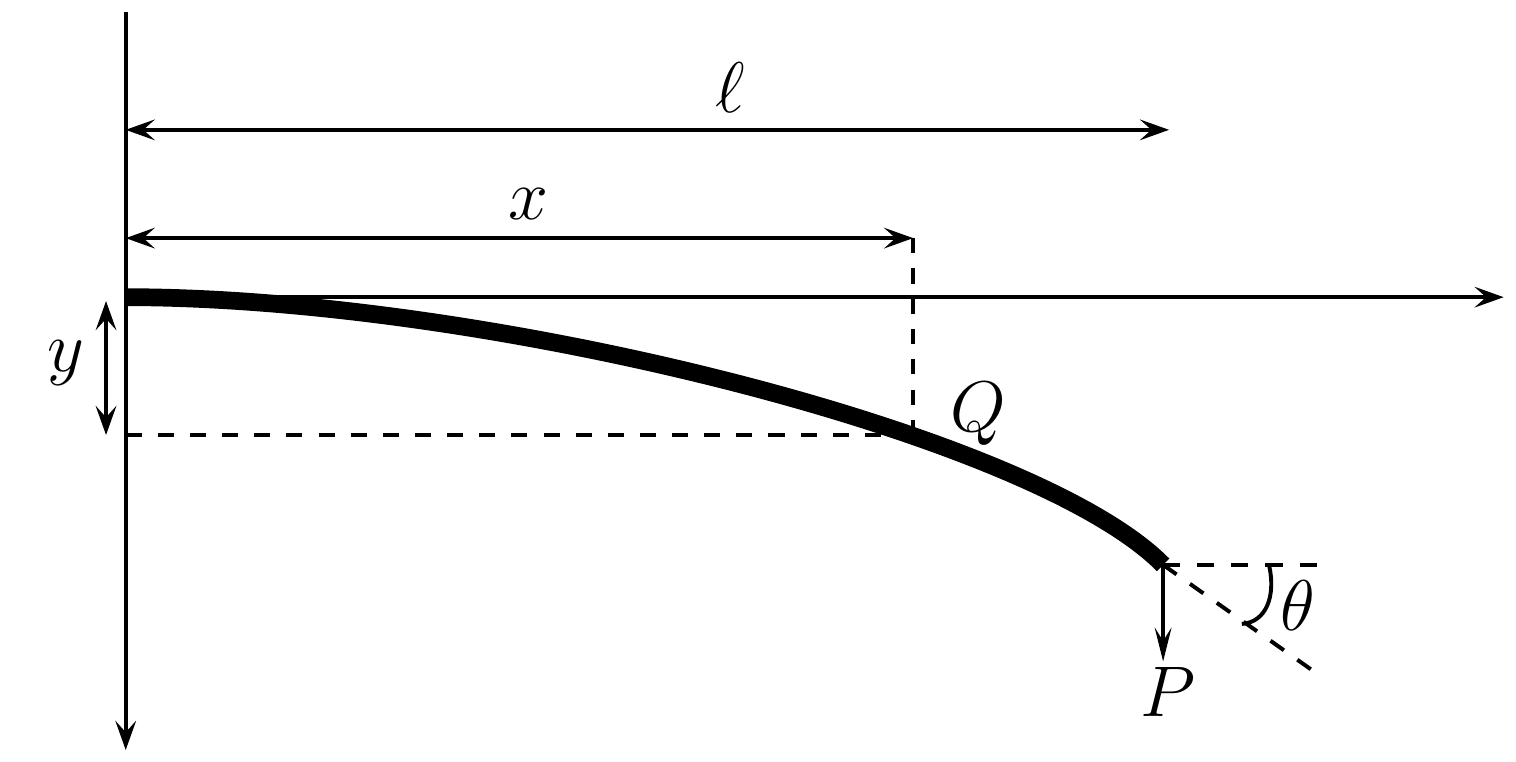}
        \captionof{figure}{Curvature of a cantilever-loaded-at-the-end of longitude $s_0$, where $\ell$ is the $x$ coordinate of the free end of the cantilever, $P$ is the load hanging from that end, and $(x,y)$ are the coordinates of a point $Q$ on the cantilever.}\label{curva}
\end{figure}

Consider a horizontal beam of length $s_0$ and uniform cross section, anchored at only one end and with a load $P=mg$ (where $m$ is the mass used) hanging from the other one, as shown in Figure~\ref{curva}. An existing model which describes the behaviour of the beam using a small slopes approximation \cite{Landau, Timoshenko_elasticity} will be shown, as well as the the dependence of the vertical displacement on the load on the pasta.

In this analysis it is assumed that the beam is made of an elastic material (one that obeys Hooke's law) and that its total length is constant \cite{Landau}. The beam's bending moment $M$ can be written as \cite{Timoshenko}
\begin{align}
\frac{d\theta}{ds} = \frac{1}{\rho}&=\frac{M}{EI},\label{flector}
\end{align}
with $\theta$ is the rotation angle of the bent curve, $\rho$ is the beam's radius of curvature, $I$ is the second moment of inertia of its cross section with respect to the rotation axis and $E$ is the Young's modulus of the material.

When the beam is bent, its cross section rotates with respect to an axis perpendicular to its neutral axis (see figure \ref{línea neutra}, this perpendicular axis would be pointing perpendicular to the paper). The cross section's second moment of inertia $I = \int_{S}r^2 dA$ can be calculated with respect to the neutral axis, where $S$ is the cross section area, $r$ is the distance from a point $P$ to the axis and $dA$ is the surface element.

Since it is possible to assign to each point in $x$ a unique $y$ coordinate, because we have $y=y(x)$, then the radius of curvature can be written as follows
\begin{align}
\rho&=\frac{\left[1+\left(\dfrac{dy}{dx}\right)^2\right]^{3/2}}{\dfrac{d^2y}{dx^2}},\label{radio curv}
\end{align}
which allows to rewrite the equation \eqref{flector} as:
\begin{align}
\dfrac{\dfrac{d^2y}{dx^2}}{\left[1+\left( \dfrac{dy}{dx}\right) ^2\right]^{3/2}}=\dfrac{M}{EI},\label{eqn bending moment and rectangular coordinates}
\end{align}
Chen \cite{Chen} defined $z=\dfrac{dy}{dx}$, so that equation \eqref{eqn bending moment and rectangular coordinates} can be written as
\begin{align}
\dfrac{dz}{dx}&=\dfrac{M(x)}{EI}(1+z^2)^{3/2}\\
\Longrightarrow \dfrac{z}{\sqrt{1+z^2}}&=\int\limits_0^x\dfrac{M(\tilde{x})}{EI}d \tilde{x}\equiv G(x).\label{g(x)}
\end{align}
Since the arc length $s$ on the beam satisfies
\begin{align}
ds&=\sqrt{1+z^2}dx,
\end{align}
the equation \eqref{g(x)} can be written as  
\begin{align}
\dfrac{ds}{dx}=\dfrac{1}{\sqrt{1-G^2(x)}}.\label{ds/dx}
\end{align}
\subsection{Small slopes analysis}
If the displacement of any given point in the loaded beam is small compared to its position when the beam is unloaded, it can be assumed that the term $(dy/dx)^2$ is negligible. This is called the \textit{small slopes approximation}:
\begin{align}
\rho&=\left(\frac{d^2y}{dx^2}\right)^{-1},\label{radio curv peq}
\end{align}
which is valid as long as 
\begin{align}
\left(\frac{dy}{dx}\right)^2& \ll 1.\label{condicion}
\end{align}
Equations \eqref{flector} and \eqref{radio curv peq} imply:
\begin{align}
\frac{d^2y}{dx^2}&=\frac{M}{EI}.\label{dif}
\end{align}
The bending moment of a cross section which lies at a distance $x$ of the origin due to a load $P$ on the free end of the beam is \cite{Timoshenko}
\begin{align}
M&=P(\ell-x).\label{m flector}
\end{align}

Replacing \eqref{m flector} in \eqref{dif} the following differential equation is obtained:
\begin{align}
\frac{d^2y}{dx^2}&=\frac{P(\ell-x)}{EI}.\label{diferential}
\end{align}

Using the initial $y(0)=0$ and boundary $\left(dy/dx\right)_{x=0}=0$ conditions, it is found that the beam's displacement curve's equation is:

\begin{align}
y(x)&=\frac{P}{6EI}[3\ell x^2-x^3].\label{pequeñaspendientes}
\end{align}

Furthermore, from equation \eqref{m flector} it is possible to verify that
\begin{align}
    G(x) = \frac{P}{2EI}(2\ell x - x^2). \label{def g(x)}
\end{align}

Note that equation~\eqref{pequeñaspendientes} applied to the condition in~\eqref{condicion} implies that
\begin{align}
    \left(\frac{P}{2EI}\right)^2(2\ell x - x^2)^2 \ll 1. \label{condicion v2}
\end{align}

Since the left hand side of equation~\eqref{condicion v2} reaches its maximum precisely in the beam's free end, where $x=\ell$, then we can conclude
\begin{align}
    \left(\frac{P\ell^2}{2EI}\right)^2 \ll 1. \label{condicion v3}
\end{align}

Solving equation~\eqref{pequeñaspendientes} for $EI$ when $x=\ell$ so that we obtain the small slopes condition \cite{Amigo_de_fajardo}

\begin{align}
    s.s. \equiv \frac{9y(\ell)^2}{4\ell^2} \ll 1. \label{condicion pp}
\end{align}

\subsection{Relation between deflection and load}
It's possible to evaluate equation~\eqref{pequeñaspendientes} at the free end of the pasta in order to obtain the relation between displacement in the $y$ axis of the beam's free end (deflection) and the load on the beam. Then, using coordinates ($x_f,y_f$)
\begin{align}
y_f&=\frac{P}{6EI}[3\ell x_f^2-x_f^3].
\end{align}
Note that in this case $x_f=\ell$ (see Figure~\ref{curva}) and we conclude
\begin{align}
y_f&=\frac{P\ell^3}{3EI}.\label{flechacarga}
\end{align}
In equation~\eqref{flechacarga} the value $\ell$ is clearly a function of the applied force, since this is the parameter that determines the deformation of the beam. However, the small slopes approximation assumes that $\ell \approx s_0$. Then, equation~\eqref{flechacarga} can be rewritten as
\begin{align}
y_f&\approx\frac{Ps_0^3}{3EI}.\label{flechacarga v2}
\end{align}
Therefore, under the small slopes approximation, the deflection $y_f$ is a function of the applied load $P$, and the slope of its graph gives the Young's modulus $E$.

\section{Experimental Setup}
\begin{figure}[h!]
    \centering
    \includegraphics[width=0.40\textwidth]{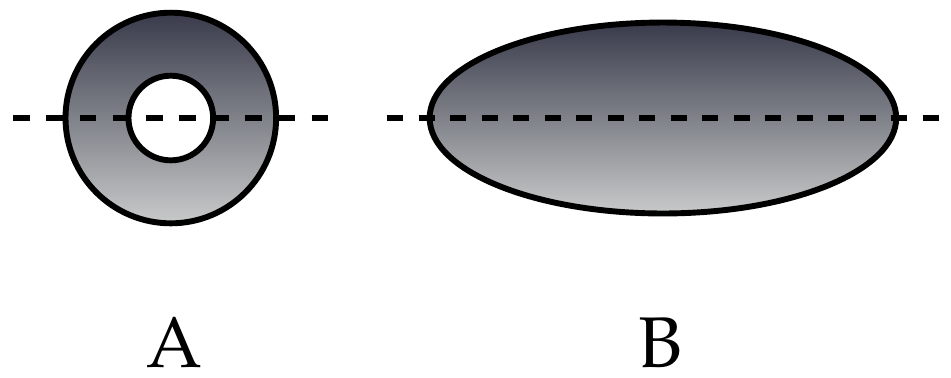}
        \captionof{figure}{Cross sections of A) bucatini and B) noodle, where the dotted lines represent the rotation axes for each pasta. Bucatini has a ring-like cross section, and the noodle has an elliptical cross section.}\label{secciones}
\end{figure}

Two different types of pasta\footnote{Colombian pasta brand Doria was used.} were used (noodles and bucatini) with different cross sections, as shown in Figure~\ref{secciones}. Pasta's lengths were measured with a common ruler, with a precision of 1 \si{\milli\metre}. The major and minor axes of the elliptical cross section of the noodle, and the inner and outer radius of the ring-like cross section of the bucatini were measured with a vernier capiler of precision 0.02 \si{\milli\metre}.

\begin{figure}[h!]
    \centering
    \includegraphics[width=0.40\textwidth]{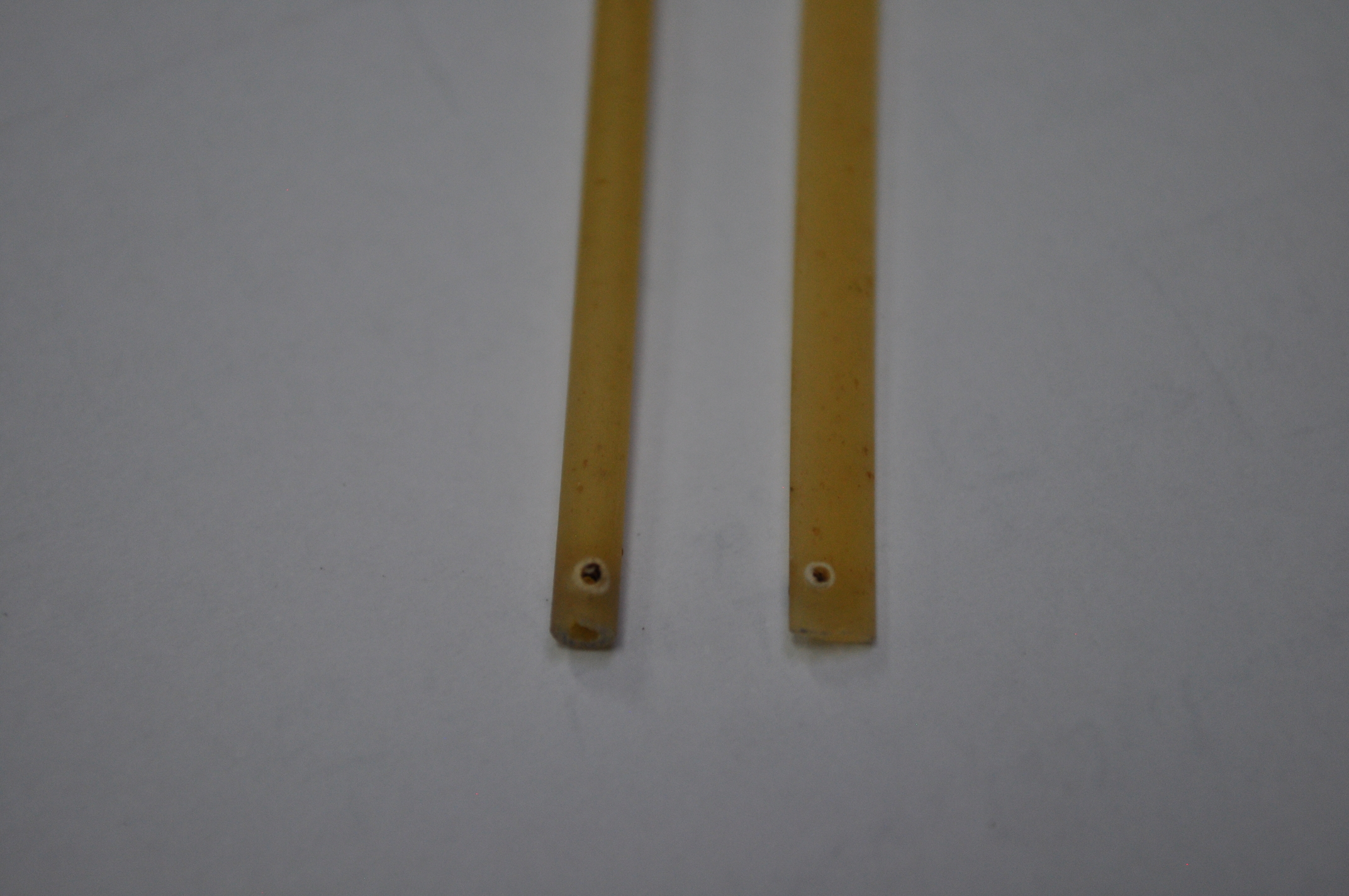}
        \captionof{figure}{Photo of a pair of bucatini and noodle, drilled at their free ends.}\label{perforadas}
\end{figure}

As it is seen in Figure~\ref{perforadas}, each pasta was drilled on one of its ends with a thin hot wire with the purpose of hanging a hook that acted as a load support, so that the loads could be easily hung from the hook. The pastas were never broken at the drilling points under load, and thus we consider that the drillings do not affect the measurements.

\begin{figure}[hb!]
    \centering
    \includegraphics[width=0.47\textwidth, clip=true, trim = 0.15cm 0 0.15cm 0]{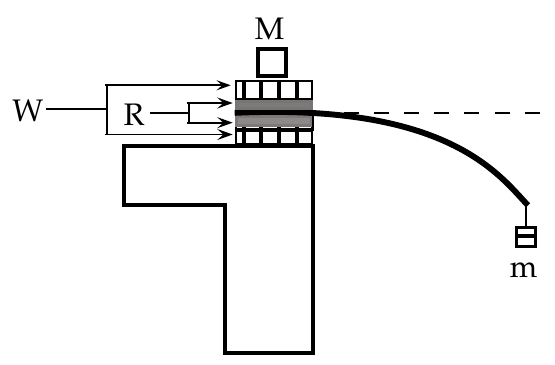}
        \captionof{figure}{Scheme of the experimental setup used for the study of the pasta's elastic properties, where $m$ is the applied load, $M$ is a mass of 3.75 \si{\kilogram}, $W$ are two flat wooden blocks, and $R$ are two hard rubber sheets.}\label{montaje}
\end{figure}

Each pasta was clamped by putting 5 \si{\centi\metre} of them between a pair of hard rubber rectangular sheets (R). These were also put between two wooden blocks (W) as recommended in \cite{brittle}. On top of the upper wooden block, a mass of 3.75 \si{\kilogram} was located in order to firmly fix the clamped end of the pasta, as shown in the Figure~\ref{montaje}. The hard rubber sheets provide firmness and avoid fractures or slips in the clamped region of the pasta.

\begin{figure}[h!]
    \centering
    \includegraphics[width=0.3\textwidth]{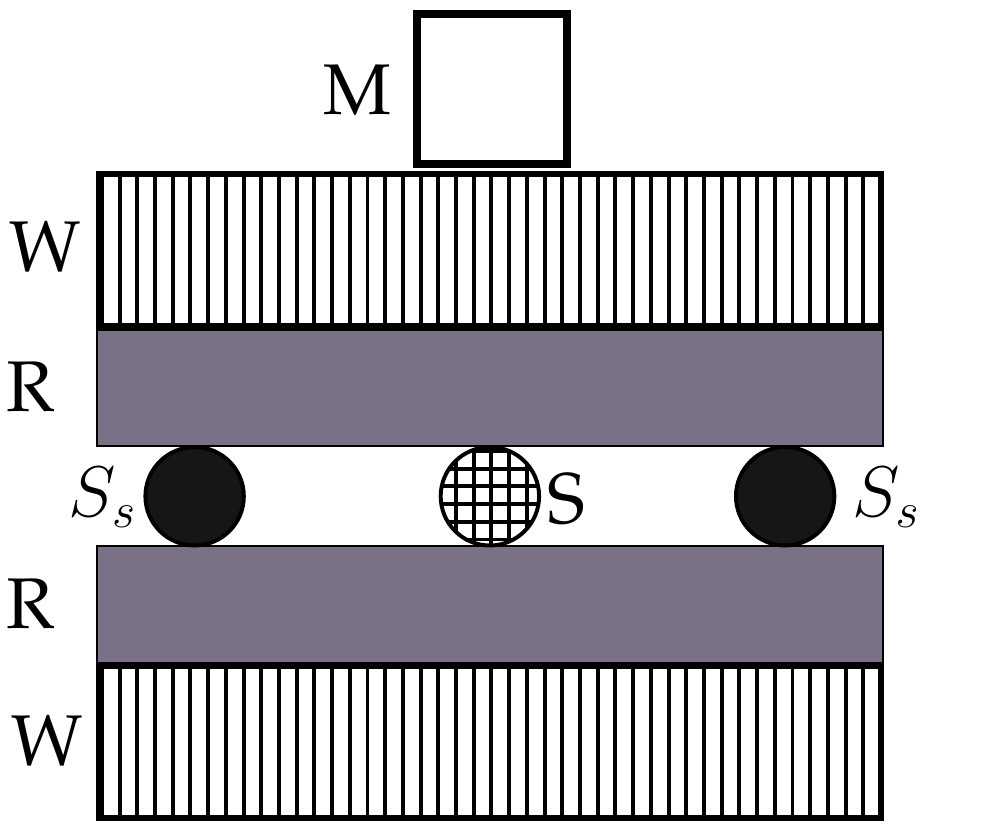}
        \captionof{figure}{Transverse view of the experimental setup shown in Figure~\ref{montaje}. $M, W$ and $R$ are the same as in Figure~\ref{montaje}, $S_s$ are a couple of pasta pieces used as lateral supports, and $S$ is the pasta to study.}\label{montaje transversal}
\end{figure}
Additionally, as seen in Figure~\ref{montaje transversal}, two pieces of pasta were located between the rubber sheets, parallel to the main pasta in order to provide stability to the upper block.

\begin{figure}[h!]
    \centering
    \includegraphics[width=0.45\textwidth]{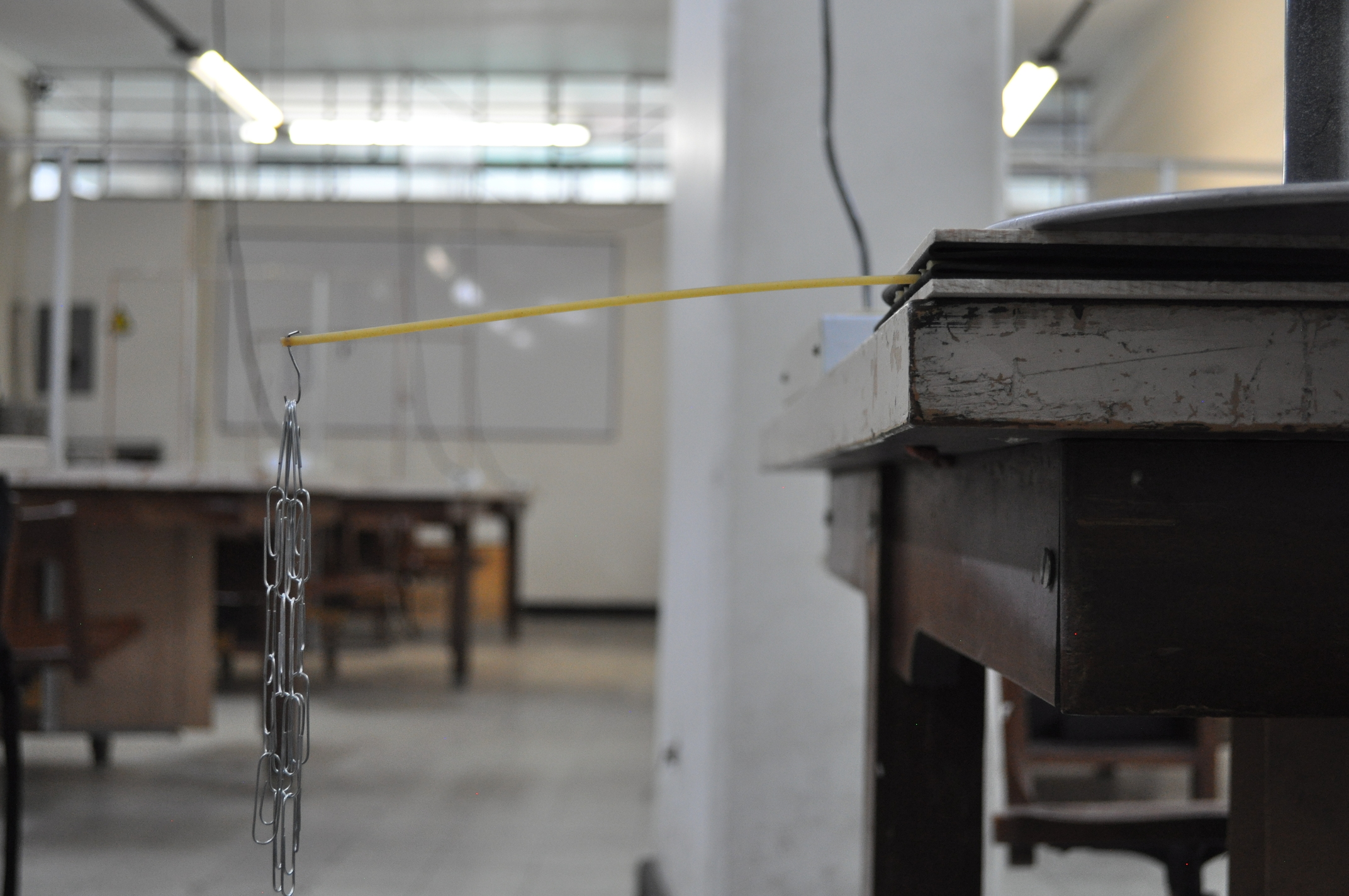}
    \caption{Picture of the cantilever setup.}
    \label{montaje foto}
\end{figure}

Metallic clips were used as loads. 100 clips were measured with a digital scale (OHAUS Scout Pro - 2000\si{\gram}) to obtain the mean value of the mass of a clip. The result was $m=$\num{0.343}$\pm$\num{1d-3}\ \si{\gram}, and we assume that all the clips weight this value. To measure the pasta's deflection, a camera was positioned at a distance of 1.5 \si{\meter} from the pasta so that the camera pointed perpendicularly to the deformation plane of the pasta. Pictures were taken while the hook was being loaded (loading process) and unloaded (unloading process). The loads (clips) were added to the hook, in intervals $\Delta t$ of 30 seconds, with care as to avoid oscillations. The software \textit{Tracker Video Analysis} \cite{Tracker} was used to process the digital images in order to measure with precision the displacements of points on the pasta. A picture of the experimental setup is shown in Figure~\ref{montaje foto}.

Three experiments were carried out. For the first one, with the pasta in distension, clips were added to the mass support in constant temporal intervals $\Delta t$. The load was added successively until the pasta broke. For each added load, the free end's deflection was measured. The second experiment consisted in loading the pasta (in the same way as in the first experiment) up to a point in which the load was less than the needed to break the pasta, then the pasta was unloaded in the same fashion with the purpose of measuring mechanical hysteresis. The behaviour of the deflection was measured during loading and unloading processes. In the third experiment, the displacement of some points marked on the pasta were measured for some fixed loads, with the purpose of fitting the elastic curve. The results of these experiments are shown in the next section.

\section{Results and discussion}\label{resultados}

\subsection{Vertical displacement vs load} \label{sec flecha vs carga}
\begin{figure}[h!]
    \centering
     \begin{tabular}{@{}p{0.45\textwidth}}
        \subfigimg[width=0.47\textwidth]{\huge{A)}}{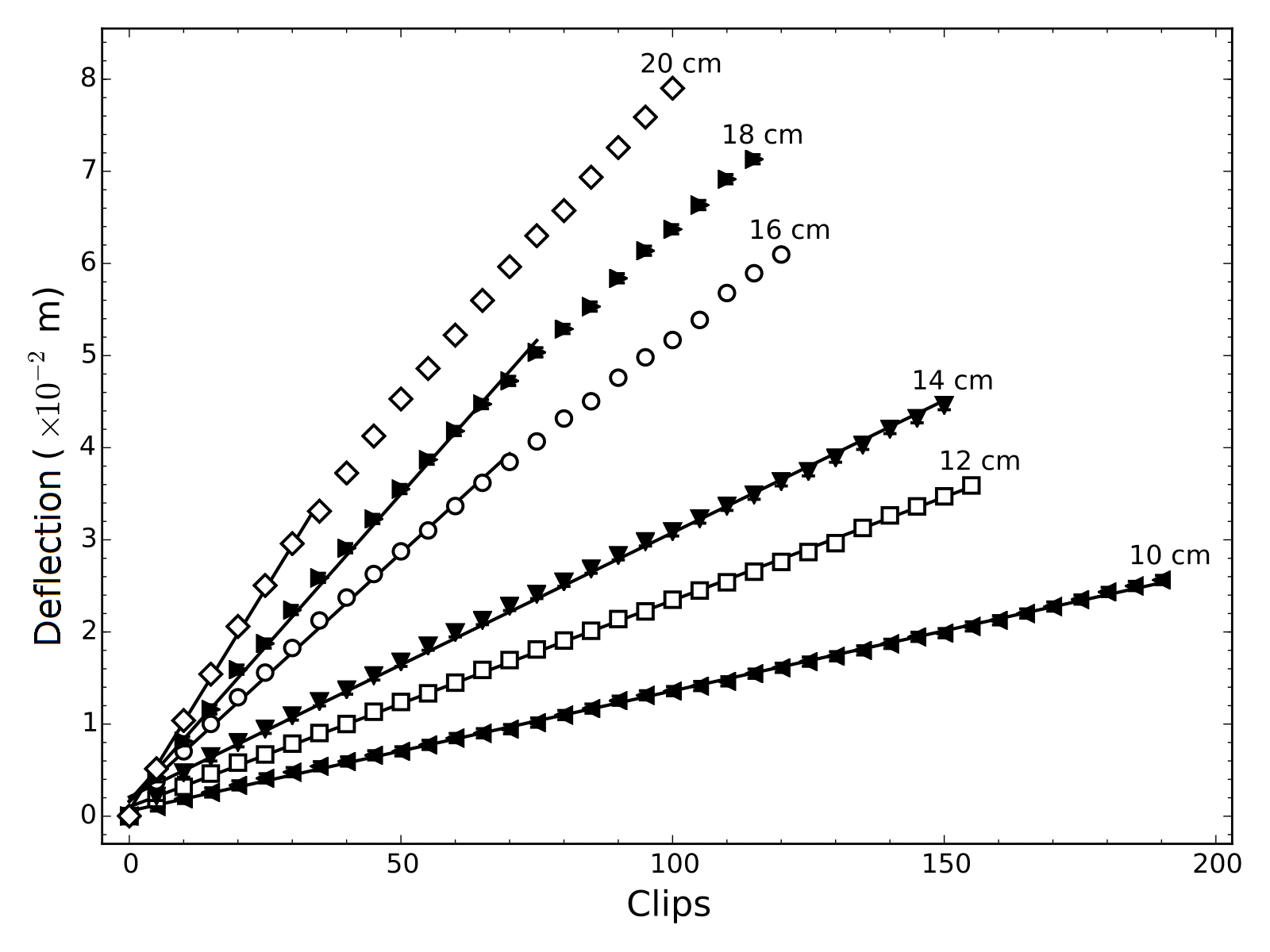}\\
        \subfigimg[width=0.47\textwidth]{\huge{B)}}{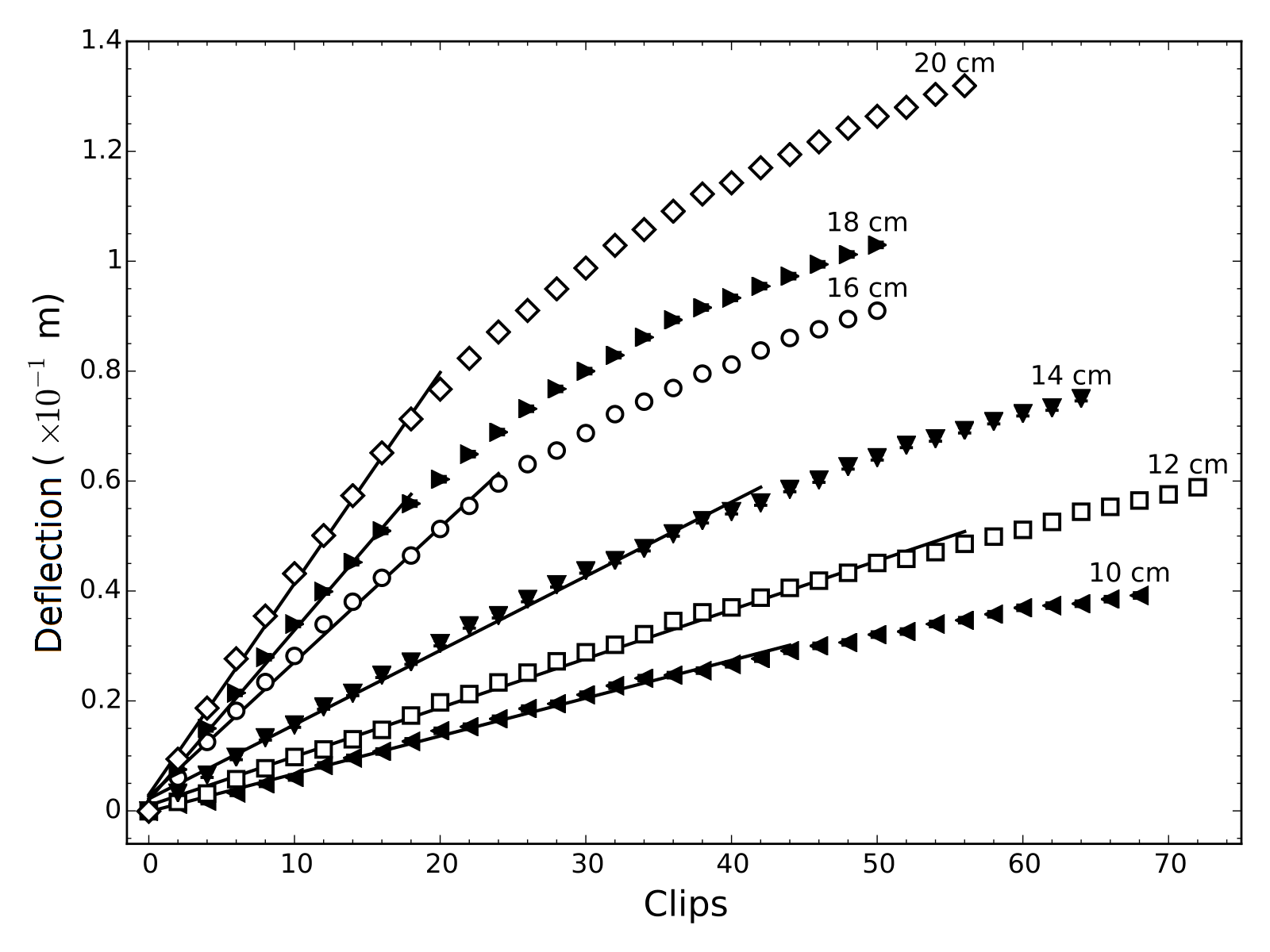}
    \end{tabular}
    \caption{Relation between vertical displacement and load for A)bucatini and B)noodles as a function of the pasta's length. Linear regressions for each set of data are shown as solid lines. The last data point of each set corresponds to the maximum load before breaking the pasta.}
    \label{grafica flechacarga pastas}
\end{figure}

Figure~\ref{grafica flechacarga pastas} shows the experimental data of vertical deflection versus load applied for 6 bucatini and 6 noodles of different lengths and the linear regression calculated with equation~\eqref{flechacarga v2}. The load was modified in intervals of $\Delta m_b = 5m$ for bucatini and $\Delta m_n = 2m$ for noodles. The data set used for the linear regression (equation~\eqref{flechacarga v2}) was selected taking into account the maximum number of experimental points whose Pearson's coefficient was greater than 0.995 for both bucatini and noodles. This criterion was used to study only the experimental data where the pastas had a linear response to the applied external force. The relation studied is not linear over the whole domain of the load, but only over the data where the loads are small. This non-linearity is more visible in the noodles as compared to the bucatini. It is also noticeable that for the same load the bucatini have a lesser vertical deflection than the noodles, this is due to the difference between the pasta's structure (see Figure~\ref{secciones}).

The values of Young's modulus found for bucatini using the slopes of Figure~\ref{grafica flechacarga pastas} are shown in Table~\ref{tabla flechacarga bucatini}. Having Young's modulus $E$ for each pasta, it is possible to solve equation~\eqref{ds/dx} to obtain the length $s$ associated to the measurement of $E$. This means that the relation $s(E)\approx s_0$ determines the validity of the approximation: if the difference between $s$ and $s_0$ is small, then the approximation is valid. The average Young's modulus for bucatini is $E=2.33\si{\giga\pascal}$.

\begin{table}
\centering
\begin{tabular}{| c | c | c |}
\hline
$s_0$ (\si{\cm})  & $E$ (\si{\giga\pascal}) & $s$(\si{\centi\metre}) \\
\hline
10 & 2.18 & 10.0  \\
12 & 2.31& 11.9  \\
14 & 2.36 & 14.0\\
16 & 2.26 & 15.8\\
18 & 2.63& 18.0\\
20 & 2.24& 20.0\\
\hline
\end{tabular}
\captionof{table}{Young's modulus ($E$) associated to the length of each bucatini using the linearity between vertical displacement and load (equation~\eqref{flechacarga v2}). Each $E$ solves equation~\eqref{ds/dx}, resulting in a value of $s$, which is the calculated length of each pasta according to the model.} \label{tabla flechacarga bucatini}
\end{table}


It can be seen in Figure~\ref{grafica flechacarga pastas} that for the noodles the vertical displacement-load relation is definitely non-linear for big loads. There are two main causes for this: the first one is the difference between the $y$ axis displacement and the real displacement of the noodle. With small deflections on the pasta the displacement is almost completely on the $y$ axis (the one that was measured). It is clear from Figure~\ref{grafica flechacarga pastas} that the pasta has a linear behaviour when the loads (and therefore the deflections) are small.  But when the displacement is big there is an important portion of it corresponding to displacement on the $x$ axis, causing a difference between the real displacement of the pasta and the one that was measured. The second cause, which is thoroughly studied in section~\ref{histéresis}, is that big loads can permanently modify the molecular arrangement of the pasta, causing not only an elastic behaviour, but also a plastic one, slightly deforming in a permanent way the length of the pasta \cite{Histeresis}.

The values found for Young's modulus using the vertical displacement versus load relation (equation~\eqref{flechacarga v2}) for noodles are shown in Table~\ref{tabla flechacarga tallarin}. The average Young's modulus for noodles is $E = 2.96$ \si{\giga\pascal}.

\begin{table}
\centering
\begin{tabular}{| c | c | c |}
\hline
$s_0$ (\si{\cm})  & $E$ (\si{\giga\pascal}) & $s$(\si{\centi\metre} )\\
\hline
10 &     2.44  &    10.1  \\
12 &     2.79 &    11.8  \\
14 &     2.92 &    13.7 \\
16 &     2.98  &    15.5 \\
18 &     3.24 &    17.3\\
20 &     3.43 &    19.0 \\
\hline
\end{tabular}
\captionof{table}{Same as Table~\ref{tabla flechacarga bucatini}, but for noodles.} \label{tabla flechacarga tallarin}
\end{table}

From tables~\ref{tabla flechacarga tallarin}~and~\ref{tabla flechacarga bucatini} the length $s$ of the pastas has a maximum discrepancy of 5\% with the measured value $s_0$. This kind of difference can be due to the used approximations \cite{Amigo_de_fajardo}. Also note from the tables that as pasta's length increases, so does the Young's modulus as well as the difference between $s$ and $s_0$. We believe this happens because, given a fixed load, the longer the pasta, the larger the deflection. This results in larger deviations from the small slopes approximation.

\subsection{Mechanical hysteresis}\label{histéresis}

If there are no external forces on the pasta it is in macroscopic equilibrium.  When external forces act upon it, its molecular arrangements oppose these forces resulting in a restoring force. This restoring force is, generally, non-conservative and it depends on the magnitude of the external force. The presence of a non-conservative force is evident when the loading and unloading processes are carried out, since it is observed the presence of the mechanical hysteresis phenomenon \cite{Histeresis}. This phenomenon consists in the restoring force being larger in the loading process than in the unloading process, resulting in a permanent deformation of the material at the end of a load-unload cycle.

To measure the mechanical hysteresis curve, a maximum load of 95 clips was hung at the free end of a bucatini, and a maximum load of 48 clips was hung at the free end of a noodle. Both pasta were 21 \si{\centi\metre} long.

\begin{figure}[h!]
    \centering
     \begin{tabular}{@{}p{0.45\textwidth}}
        \subfigimg[width=0.47\textwidth]{\Huge{A)}}{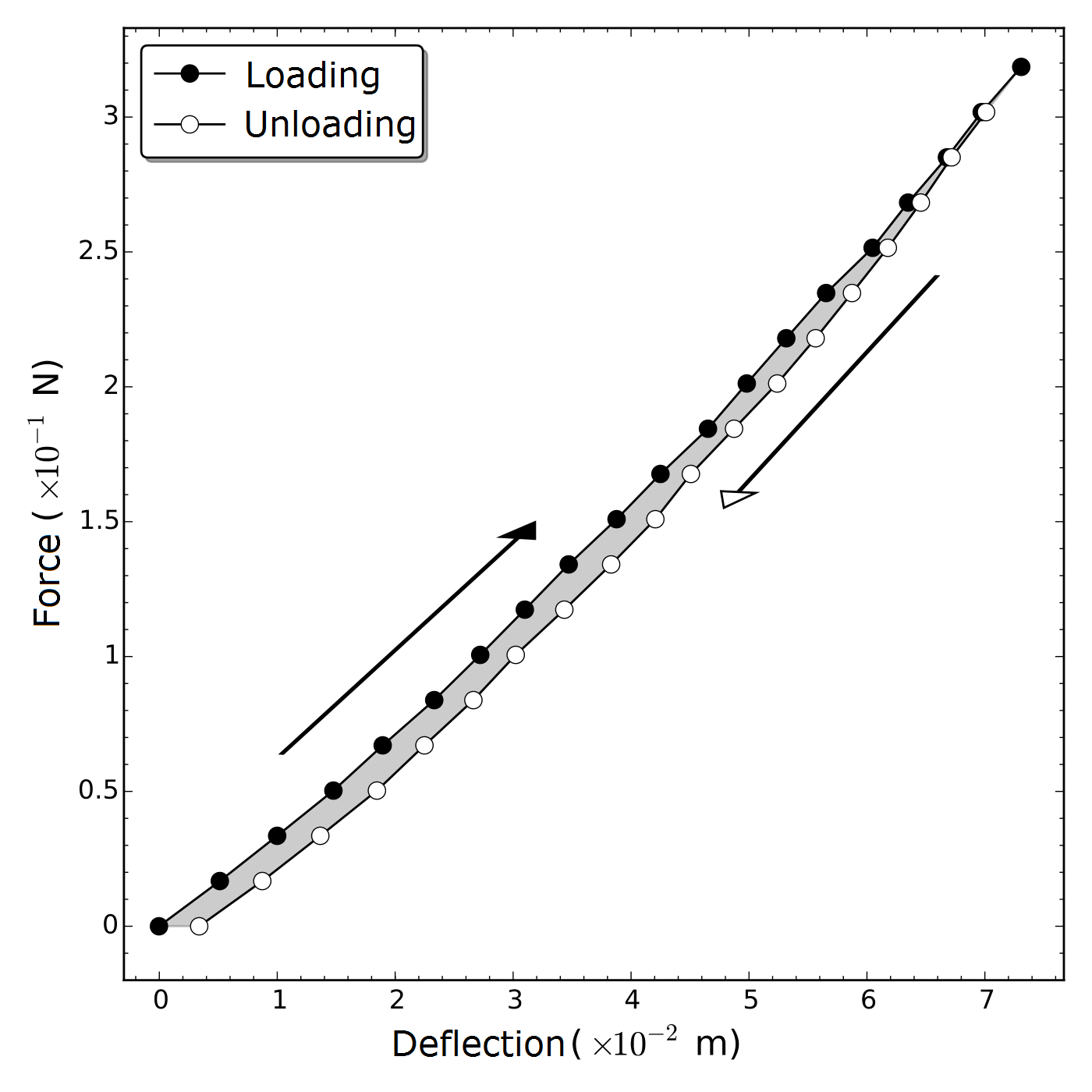}\\
        \subfigimg[width=0.47\textwidth]{\Huge{B)}}{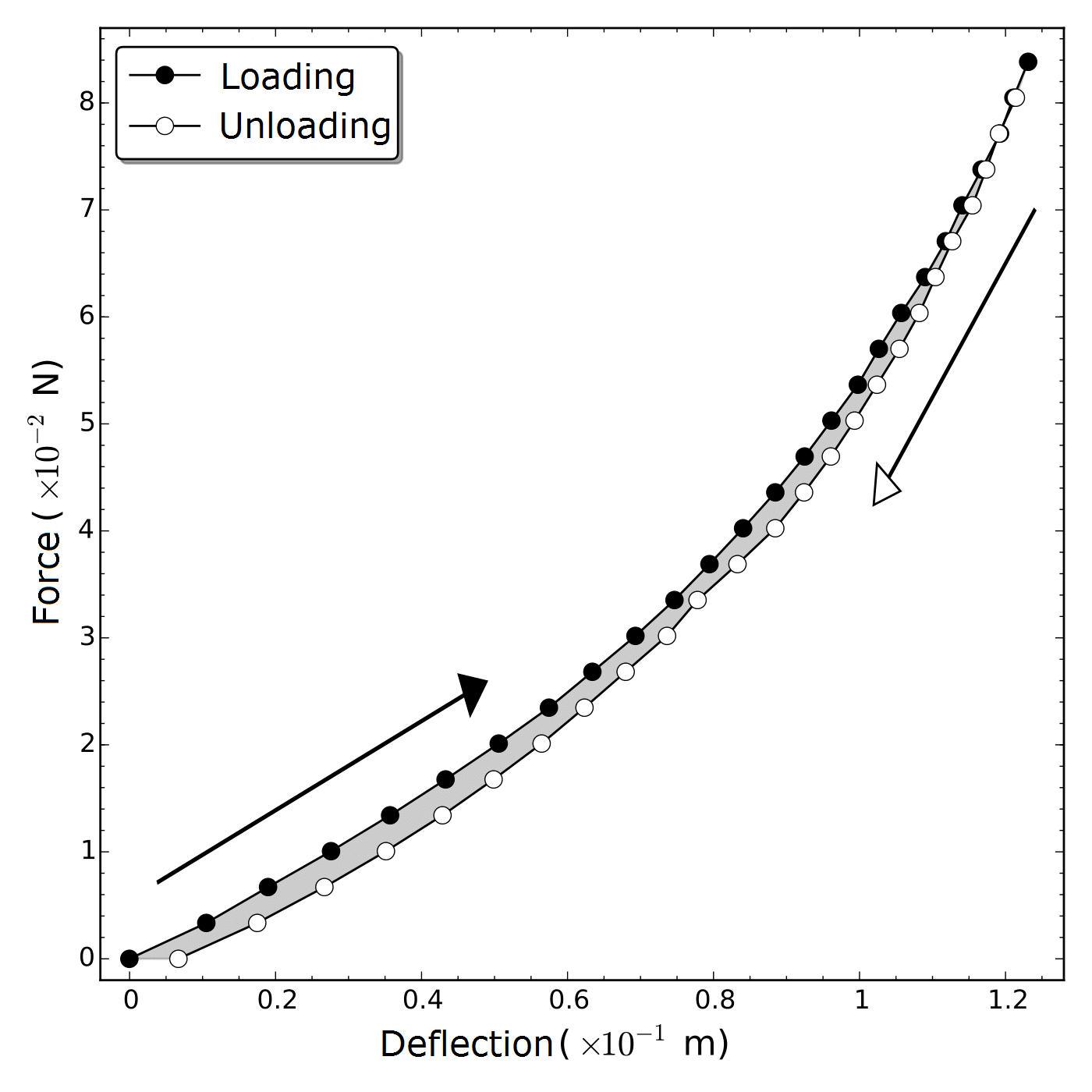}
    \end{tabular}
    \caption{Hysteresis curves for A) bucatini and B) noodle, each one with a length of 21 \si{\centi\metre}. The black dots represent experimental measurements corresponding to the loading process and the white dots the ones corresponding to the unloading process. The arrows show the temporal order of the measurements.}
    \label{histeresis}
\end{figure}

Both bucatini and noodles have small hysteresis, being the loading curve slightly different from the unloading curve. The shaded area in Figures~\ref{histeresis} represents the work done by non-conservative forces. This work is related to the change in the pasta's curvature. The work for the bucatini is $W=8\times10^{-4}\si{\joule}$ and for the noodle is $W=3.14\times10^{-4}\si{\joule}$. We conclude that the pasta present mechanical hysteresis, resulting in a small elongation and curvature after the loading-unloading cycle.

\subsection{Elastic curve} \label{curva elástica}
\begin{figure}[h!]
    \centering
     \begin{tabular}{@{}p{0.45\textwidth}}
        \subfigimg[width=0.47\textwidth]{\huge{A)}}{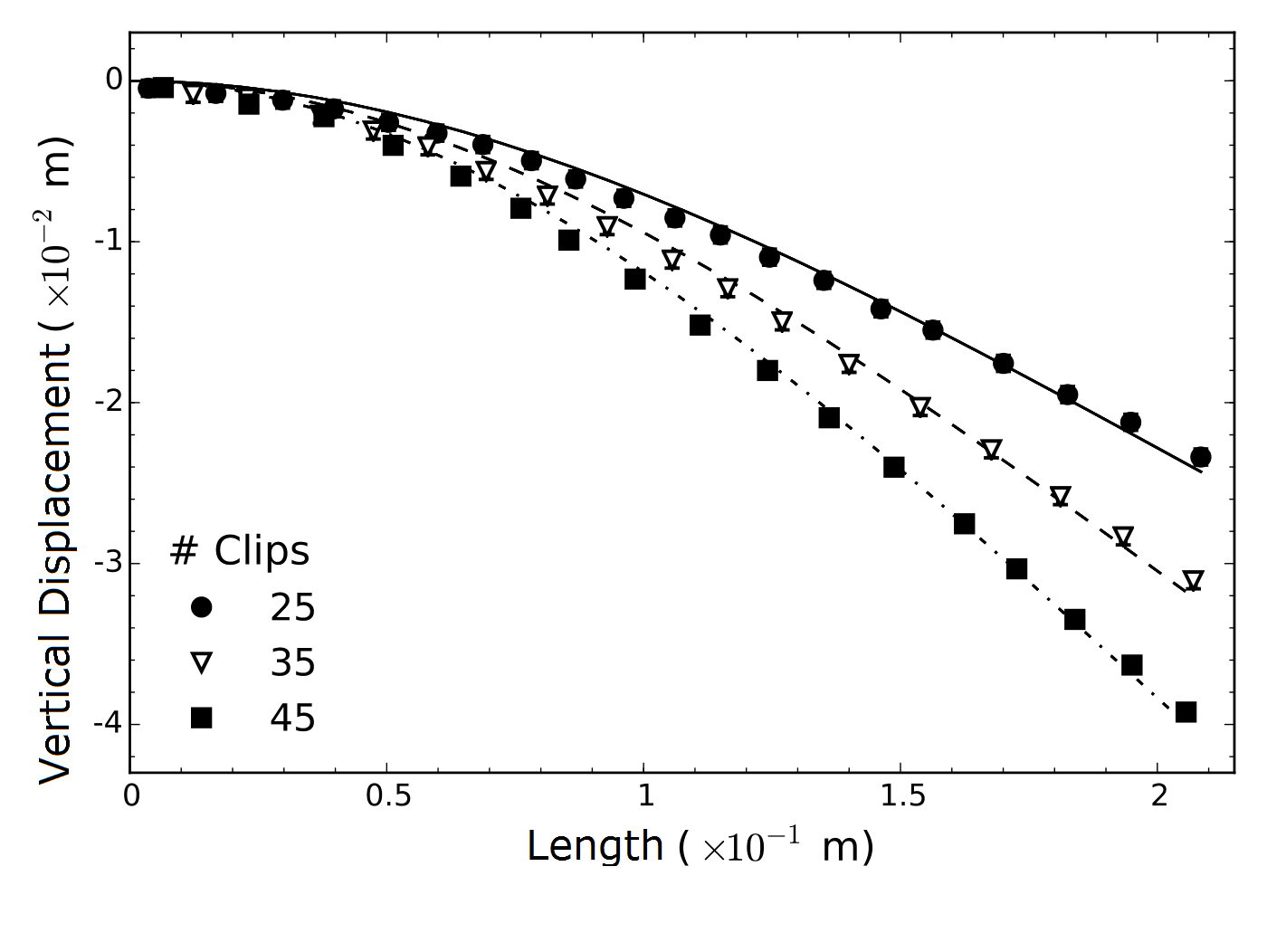}\\
        \subfigimg[width=0.47\textwidth]{\huge{B)}}{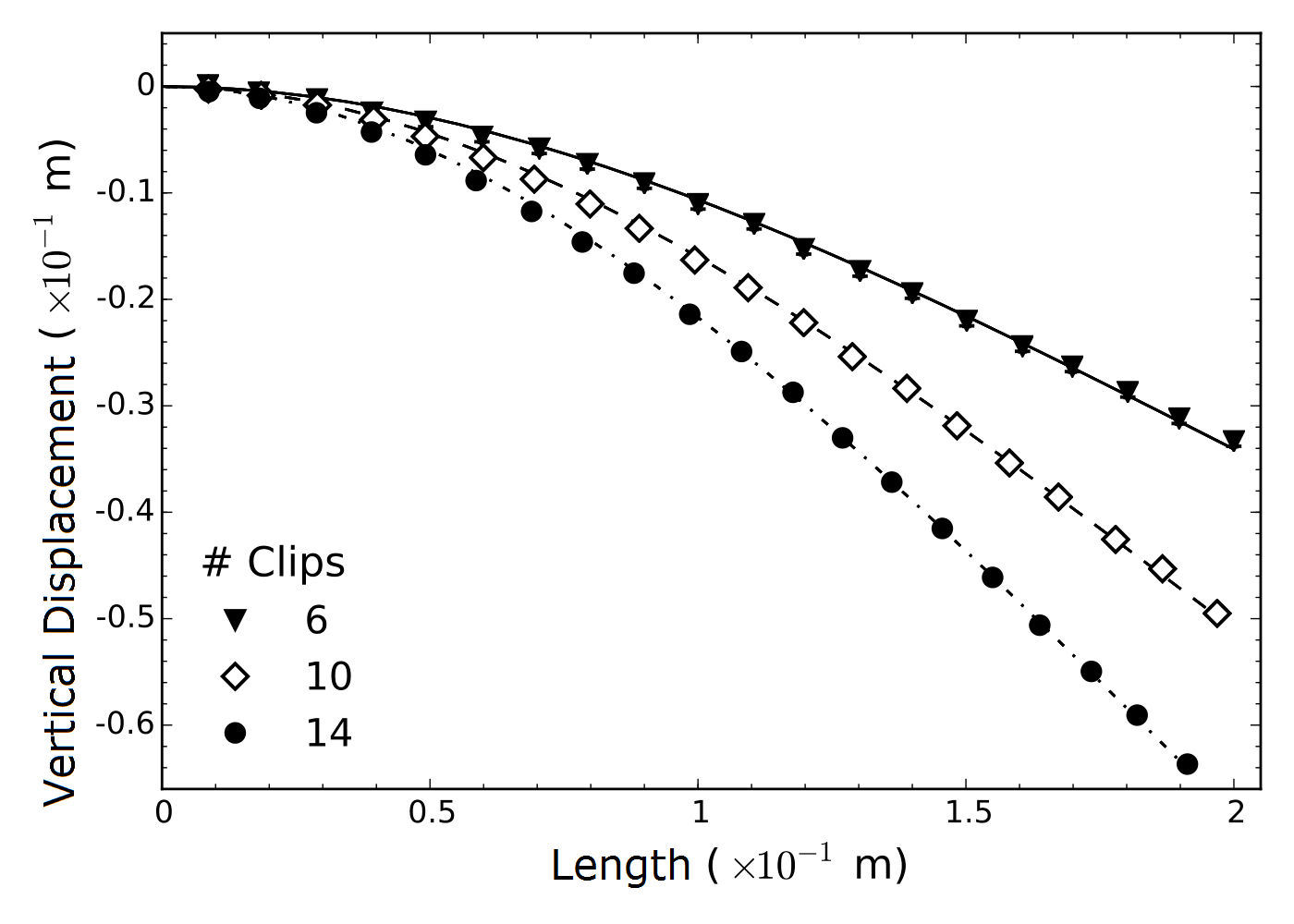}
    \end{tabular}
    \caption{Experimental data of beam's curvature for A)  bucatini of 21 \si{\centi\metre} and B) noodle of 20 \si{\centi\metre}, subjected to three different loads (clips). The elastic curves were found fitting equation~\eqref{pequeñaspendientes}.}
    \label{pastas mal}
\end{figure}

The small slopes equation~\eqref{pequeñaspendientes} was fitted to a set of points distributed onto the pasta in order to find the elastic curves of a 21 \si{\centi\metre} bucatini and a 20 \si{\centi\metre} noodle, subjected to three different fixed loads. Figures~\ref{pastas mal} show the experimental data, as well as fitting curves for both pasta. To satisfy condition~\eqref{condicion pp}, three small loads were selected so that the deformation in the pasta was not too large.

Since each curve in Figures~\ref{pastas mal} is a function of Young's modulus, then the small slopes method allows us to find the modulus that better fits the data. Young's moduli associated to bucatini curves are shown in Table~\ref{tabla pp bucatini}.
\begin{table}
\centering
\begin{tabular}{|c|c|c|c|}
\hline
Clips & $E$ (\si{\giga\pascal}) & $s.s. \times 10^{-1}$ & $s(\si{\centi\metre})$\\
\hline
25 & 2.53   &   0.28$\pm$0.01   &   21.0\\
35 & 2.62   &   0.51$\pm$0.02   &   21.0 \\
45 & 2.65   &   0.82$\pm$0.02   &   21.1\\
\hline
\end{tabular}
\captionof{table}{Young's modulus $E$ associated to the loads (units are clips) for the 21 \si{\centi\metre} bucatini using the small slopes method (equation~\eqref{pequeñaspendientes}). The $s.s.$ coefficient is also shown, as well as the length $s$ associated to each $E$ obtained with the equation~\eqref{ds/dx}.}\label{tabla pp bucatini}
\end{table}

From Table~\ref{tabla pp bucatini} a mean value of $E=2.60\ \si{\giga\pascal}$ is obtained for the Young's modulus for the 21 \si{\centi\metre} bucatini. This value is larger compared to the one found in section~\ref{sec flecha vs carga} (a difference of 10\%). This occurs because the equation~\eqref{pequeñaspendientes} of the elastic curve does not take into account the $\ell\approx s_0$ approximation made in section~\ref{sec flecha vs carga}.

Note that the length values computed for the different loads shown in Table~\ref{tabla pp bucatini} coincide with the 21 \si{\centi\metre} bucatini's length. This shows that the small slopes approximation is valid. In the case of noodles, the values of Young's modulus found with equation~\eqref{pequeñaspendientes} are shown in Table~\ref{tabla pp tallarin}.

\begin{table}
\centering
\begin{tabular}{|c|c|c|c|}
\hline
Clips & $E$ (\si{\giga\pascal}) & $s.s. \times 10^{-1}$ & $s$(\si{\centi\metre})\\
\hline
6 & 2.15   &   0.67  &   19.8 \\
10 & 2.32    &   1.47  &   20.0  \\
14 & 2.32   &   2.59  &   20.2 \\
\hline
\end{tabular}
\captionof{table}{Same as Table~\ref{tabla pp bucatini}, but for noodles.}\label{tabla pp tallarin}
\end{table}

In this case a mean value of $E = 2.26 \ \si{\giga\pascal}$ was foundrm the validity of the approximation ($\ell\approx s_0$), equation~\eqref{ds/dx} was solved. In this case, it must be true that $G^2(x) < 1$ in order to carry out a numerical integration. Since $G(x)$ reaches a maximum when $x=\ell$, then solving the equation~\eqref{ds/dx} depends on the truth value of
\begin{align}
    \frac{P\ell^2}{2EI} < 1. \label{condicion G}
\end{align}

For bucatini and for noodles, small loads \textit{must be} chosen so that the condition~\eqref{condicion G} is satisfied. Taking into account the lengths $s$ computed with the equation~\eqref{ds/dx} shown in Tables~\ref{tabla pp bucatini}~and~\ref{tabla pp tallarin}, and comparing them with the lengths shown in Tables~\ref{tabla flechacarga bucatini}~and~\ref{tabla flechacarga tallarin}, we see that the former show slightly better accuracy than the later. This suggests that the elastic curve method provides more precision than equation~\eqref{flechacarga v2} when calculating the Young's modulus.


\section{Conclusions}

The small slopes approximation allows the use of vertical displacement versus load data, to obtain the pasta's Young's modulus. This approximation is valid to determine the value of $E$ in fragile materials such as pasta, as long as small loads are used, so that there is a linear relation between vertical displacement and load. For the bucatini it was found that $E=2.33 \ \si{\giga\pascal}$ and for the noodle $E = 2.44 \ \si{\giga\pascal}$. From the elastic curve it was found that the values of Young's modulus are $E=2.60\ \si{\giga\pascal}$ for bucatini and $E=2.26\ \si{\giga\pascal}$ for noodles.

Once the pasta's Young's modulus are calculated with the elastic curve method, the theoretical length $s$ of each pasta can be found. These lengths are consistent with the measured lengths $s_0$. However, when the value of Young's modulus calculated with the small slopes method was used, discrepancies of up to 5\% were found between the computational and the measured length of the pasta. Despite this, both equations are found to be appropriated for the determination of Young's modulus in fragile and brittle materials such as pasta.

Mechanical hysteresis was found in the pasta, suffering permanent deformations after a loading-unloading cycle. The work performed by non-conservative forces that causes this deformation was found to be $W=8\times10^{-4}\  \si{\joule}$ for the bucatini and $W=3.14\times10^{-4}\ \si{\joule}$ for the noodle. The experiments described in this paper make up a simple and effective method for the study of the elastic behaviour of accessible materials for any laboratory.

\end{document}